\documentclass[11pt]{article}

\usepackage{amsmath}
\usepackage{amssymb}
\usepackage{graphics}
\usepackage{rotating}
\usepackage{cite}
\usepackage{color}
\usepackage{fancybox}
\usepackage{pstricks}

\def\0{\mbox{\tiny $0$}}
\def\1{\mbox{\tiny $1$}}
\def\2{\mbox{\tiny $2$}}
\def\3{\mbox{\tiny $3$}}
\def\4{\mbox{\tiny $4$}}
\def\5{\mbox{\tiny $5$}}
\def\6{\mbox{\tiny $6$}}
\def\7{\mbox{\tiny $7$}}
\def\8{\mbox{\tiny $8$}}
\def\9{\mbox{\tiny $9$}}
\title{\shadowbox{\large \bf THE SNELL LAW FOR  QUATERNIONIC POTENTIALS}}

\author{
\small  Stefano De Leo\thanks{Department of Applied Mathematics,
State University of Campinas, Brazil [deleo@ime.unicamp.br] } \,\,
and\, Gisele C. Ducati\thanks{CMCC,
Universidade Federal do ABC, S\~ao Paulo, Brazil [ducati@ufabc.edu.br]}}

\date{\small
\fcolorbox{black}{yellow} {\color{red} $\bullet$ {\color{black}{
{\footnotesize  {\sc Journal of Mathematical Physcis} {\bf 54},122109-9 (2013) }}}
{\color{red}{$\bullet$}} } }

\begin{document}
%
%%%%%%%%%%%%%%%%%%%%%%%%%%%%%%%% PAPER %%%%%%%%%%%%%%%%%%%%%%%%%%%%%%%%%%%%%

\maketitle

\vspace*{-.7cm}

\begin{abstract}
\noindent By using the analogy between optics and quantum mechanics, we obtain the Snell law for the planar motion of quantum particles in the presence of quaternionic potentials.
\end{abstract}

%%%%%%%%%%%%%%%%%%%%%%%%%%%%%%%%%%%%%%%%%%%%%%%%%%%%%%%%%%%%%%%%%%%%%%%
%%%%%%%%%%%%%%%%%%%%%%%%%%%%%%%%%%%%%%%%%%%%%%%%%%%%%%%%%%%%%%%%%%%%%%%

%%%%%%%%%%%%%%%%%%%%%%%%%%%%%%%%%%%%%%%%%%%%%%%%%%%%%%%%%%%%%%%%%%
%%%%%%%%%%%%%%%%%%%%%%%%%%%%%%%%%%%%%%%%%%%%%%%%%%%%%%%%%%%%%%%%%%%%%%%

%%%%%%%%%%%%%%%%%%%%%%%%%%%%%%%%%%%%%%%%%%%%%%%%%%%%%%%%%%%%%%%%%%%%%%%
%%%%%%%%%%%%%%%%%%%%%%%%%%%%%%%%%%%%%%%%%%%%%%%%%%%%%%%%%%%%%%%%%%%%%%%

%\PACS{ {??.??.?} \and  {??.??.?}{}}
%\PACS{ {42.25.Bs, 42.25.Gy, 42.50.Xa (PACS).}{}}

% Warning: No PACS code given

%02.10.Hh Rings and algebras
%02.10.Ud Linear algebra
%02.10.Yn Matrix theory

%02.30.Hq Ordinary differential equations
%02.30.Jr Partial differential equations
%02.30.Tb Operator theory

%03.65.-w Quantum mechanics
%03.65.Ca Formalism
%03.65.Ta Foundations of quantum mechanics;
%03.65.Xp Tunnelling, traversal time, quantum Zeno dynamics

%12.15.F Quarks and lepton masses and mixing
%14.60.Pq Neutrino mass and mixing

%\offprints{~Stefano De Leo.}

%%%%%%%%%%%%%%%%%%%%%%%%%%%%%%%%%%%%%%%%%%%%%%%%%%%%%%%%%%%%%%%%%%%%%%%
%%%%%%%%%%%%%%%%%%%%%%%%%%%%%%  SECTION   %%%%%%%%%%%%%%%%%%%%%%%%%%%%%
%%%%%%%%%%%%%%%%%%%%%%%%%%%%%%%%%%%%%%%%%%%%%%%%%%%%%%%%%%%%%%%%%%%%%%

\section*{\normalsize I.INTRODUCTION}

Analogies between quantum mechanics\cite{cohen,grif} and optics\cite{born,saleh} are well known.
The possibility to propose optical experiments to study the fascinating behavior of quantum mechanical system is a subject of great interest in litterature\cite{DRAG,SL}. The recent progress\cite{Q1,Q2,Q3,Q4,Q5,Q6,Q7,Q8,Q9,Q10} in looking for quantitative and qualitative differences between complex and quaternionic quantum mechanics\cite{ADL} surely have contributed to
make the subject more useful to and accessible over the worldwide community of scientists interested in testing the existence of quaternionic potentials. Stimulated by the analogy between quantum mechanics and optics, in this paper we propose a quantum mechanical study of the planar motion in the presence of a quaternionic potential which lead to a {\em new} Snell law.

In the next section, by considering the ordinary Schr\"odinger equation and complex wave functions,
we obtain  the standard Snell law by using the planar motion in the presence of complex potentials. In the section III,
 we introduce the  quaternionic Schrodinger equation and calculate, for quaternionic wave functions,
 the {\em new} Snell law.  In the following section, we obtain the reflection coefficient
for the planar motion in presence of quaternionic potentials and compare the complex case with the pure quaternionic case. Section V contains a discussion of the results, a proposal for further studies and, finally, our conclusions.

\section*{\normalsize II. COMPLEX QUANTUM MECHANICS AND SNELL LAW }

Consider an optical beam moving from a dielectric medium with refractive index $n_{\1}$ to a second dielectric medium of refractive index $n_{\2}<n_{\1}$. If the incoming beam forms an angle $\theta$
with respect to the perpendicular direction to the stratification, the beam, for incidence angle $\theta<\theta_c=\mbox{arcsin}[n_{\2}/n_{\1}]$ will be transmitted in the second medium and will be deflected forming an angle $\varphi$ given by the well know Snell law\cite{born,saleh},
\begin{equation}
\sin\theta = n\,\sin\varphi\,\,,
 \end{equation}
 where $n=n_{\2}/n_{\1}$.  For $\theta>\theta_c$ we have total reflection.

 The analogy between optics and quantum mechanics allows to obtain this fascinating law by analyzing the reflected and refracted waves for the  planar  $(y,z)$ motion  of a quantum mechanical particle in the presence of a stratified potential\cite{ste1,ste2},
\begin{equation}
V_{\1}(z_*) = \left\{\,0\,\,\,\,\,\mbox{for}\,\,\,z_*<d_*\,\,\,\,\,\,\,\mbox{and}\,\,\,\,\,\,\,
V_{\1}\,\,\,\,\,\mbox{for}\,\,\,z_*>d_*\,\right\}\,\,,
\end{equation}
whose discontinuity is at a distance $d_*$ from the particle source along the $z_*$-stratification axis forming with the incident direction $z$ an angle $\theta$ (see Fig.\,1a),
\begin{equation*}
\left( \begin{array}{c}
y_* \\
z_* \end{array} \right)= \left( \begin{array}{rr}
\cos\theta & \sin\theta \\
-\sin\theta & \cos\theta \end{array} \right)\,\, \left( \begin{array}{c}
y\\
z \end{array} \right)~.
\end{equation*}
The solution of the Schr\"odinger equation for $z_*<d_*$ (region I) is given by
\begin{equation}
\psi_{_I}(y_*,z_*) = \psi_{_{\mbox{\tiny inc}}}(y_*,z_*) + r\,\psi_{_{\mbox{\tiny ref}}}(y_*,z_*)
\end{equation}
where $r$ is the reflection amplitude\cite{cohen},
\begin{equation}
\label{rc}
r = \displaystyle \frac{p_{z_{*}} - q_{z_{*}}}{p_{z_{*}} + q_{z_{*}}}\,\,\exp[\,2\,i\,p_{z_{*}}d_*/\,\hbar\,]\,\,,
\end{equation}
$\psi_{_{\mbox{\tiny in}}}$ the incoming wave moving along the $z$-axis,
\begin{equation}
\psi_{_{\mbox{\tiny in}}}(y_*,z_*) = \exp[\,i\,p\,z\,/\hbar\,]=\exp[\,i\,(\,p_{y_{*}}\,y_{*}+ p_{z_{*}}\,z_{*})\,/\hbar\,]\,\,,\end{equation}
and, finally, $\psi_{_{\mbox{\tiny ref}}}$ the reflected wave,
\begin{equation}
\psi_{_{\mbox{\tiny ref}}}(y_*,z_*) =\exp[\,i\,(\,p_{y_{*}}\,y_{*}- p_{z_{*}}\,z_{*})\,/\hbar\,]~.
\end{equation}
Region I ($z_*<d_*$) represents the plane zone which is potential free. Consequently, from the Schr\"odinger equation, we obtain the well known momentum/energy relation,
\begin{equation*}
p_{y_{*}}^{^{2}}+\,p_{z_{*}}^{^{2}}=p^{^{2}}=2\,m\,E\,\,.
\end{equation*}
In region II ($z_*>d_*$), the solution is given by
\begin{equation}
\psi_{_{II}}(y_*,z_*) = t\,\exp[\,i\,(\,p_{y_{*}}\,y_{*}+ q_{z_{*}}\,z_{*})\,/\hbar\,]\,\,,
\end{equation}
where $t$ is the transmission amplitude and, due to the fact that the discontinuity is along $z_*$, the  momentum component perpendicular to this axis  is not influenced by the potential, consequently  $q_{y_*}=p_{y_*}$. From the Schr\"odinger equation in the presence of step-wise potentials ($z_*>d_*$),
\begin{equation}
\label{cse}
 E\,\psi_{_{II}}(y_*,z_*) = -\,\left[\,\frac{\,\,\,\hbar^{^{2}}}{2\,m}\,\left(\,\partial_{y_*y_*}+\partial_{z_*z_*}\, \right)- V_{\1}\,\right]\,\psi_{_{II}}(y_*,z_*)\,\,,
\end{equation}
we obtain the following momentum/energy relation
\begin{equation}
\label{pev1}
  p_{y_{*}}^{^{2}} + \,q_{z_{*}}^{^{2}} = 2\,m\,( E - V_{\1} ) = \left(1-\frac{V_{\1}}{E}\right)\,p^{\2}= n^{\2}p^{\2}\,\,,
\end{equation}
where we have introduced the dimensionless quantity
\begin{equation} n =  \sqrt{1-\frac{V_{\1}}{E}}~.\end{equation}
The incoming wave has a momentum $p$ and  moves along $z$. The incidence angle with respect to the $z_{*}$-axis is $\theta$,
\[ p_{z_{*}}=p\,\cos\theta\,\,.\]
The transmitted wave  moves in the potential region with  momentum $np$. The transmitted
angle is $\varphi$ (see Fig\,1a), consequently
\[ q_{z_{*}}=np\,\cos\varphi\,\,.\]
From Eq.(\ref{pev1}), we find
\begin{equation*}
p^{\2}\,\sin^{\2}\theta + n^{\2}p^{\2}\,\cos \varphi^{\2} = n^{\2} p^{\2}
\end{equation*}
which implies
\begin{equation}
\label{snell-c}
\sin \theta =n \,\sin \varphi\,\,.
\end{equation}
Thus, we recover  the Snell law given in the beginning of this section. As it is illustrated in Fig.\,1a, an incoming particle with energy $E=3\,V_{\1}$ which moves along the $z$-axis forming an angle $\pi/4$ with respect to the $z_*$-axis,  will be deflected,  in the potential region,
forming an angle
\[ \varphi = \arcsin\left[\,\frac{1}{n}\,\sin\theta\,\right] = \arcsin\left[\,\sqrt{\frac{3}{2}}\,\frac{1}{\sqrt{2}}\,\right]=\frac{\pi}{3}
\]
with respect to the $z_*$-axis. Before of concluding this section, we observe that
from the continuity equations for the wave function and its derivative at the discontinuity $z_*=d_*$, we find the following reflection amplitude
\begin{equation}
\label{rcom}
r = \frac{\cos\theta - \sqrt{n^{\2}-\sin^{\2}\theta}}{\cos\theta + \sqrt{n^{\2}-\sin^{\2}\theta}}\,\exp[\,2\,i\,p_{z_*}d_*]
  =  \frac{1- n^{\2}}{(\,\cos\theta + \sqrt{n^{\2}-\sin^{\2}\theta}\,)^{^{2}}}\,\exp[\,2\,i\,p_{z_*}d_*]\,\,.
\end{equation}
From the reflection amplitude is immediately seen that for $\sin\theta>n$,
\begin{eqnarray}
r &=& \frac{1- n^{\2}}{(\,\cos\theta + i\,\sqrt{\sin^{\2}\theta -n^{\2}}\,)^{^{2}}}\,\exp[\,2\,i\,p_{z_*}d_*] \nonumber \\& =&
\exp\left[\,2\,i\,\left(\,p_{z_*}d_*  - \mbox{arctan}\left[\frac{\sqrt{\sin^{\2}\theta -n^{\2}}}{\cos\theta}\right]\,\right)\right]\,\,,
\end{eqnarray}
we have total internal reflection. In Fig.\,1a, where $E=3\,V_{\1}$, the critical angle is given by
\begin{equation}
\theta_c=\arcsin\left[\,\sqrt{\frac{2}{3}}\,\right]\approx \frac{\pi}{3.29}\,\,.
\end{equation}

\section*{\normalsize III. QUATERNIONIC QUANTUM MECHANICS AND  SNELL LAW}

In the previous section, we have obtained the Snell law and the reflection coefficient for planar motion in the presence of complex potentials. Is the Snell law somehow modified if we
repeat the previous analysis for Schr\"odinger equation in the presence of a quaternionic potential? The answer is yes and we are going show how.

The time-independent Schr\"odinger equation in the presence of a quaternionic potential,
\[ \boldsymbol{h}\cdot \boldsymbol{V}(z_*) =\left\{\,0\,\,\,\,\,\mbox{for}\,\,\,z_*<d_*\,\,\,\,\,\,\,\mbox{and}\,\,\,\,\,\,\,
i\,V_{\1}+j\,V_{\2}+k\,V_{\3}\,\,\,\,\,\mbox{for}\,\,\,z_*>d_*\,\right\}\,\,,\]
is given by\cite{Q1,ADL}
\begin{eqnarray}
\label{qse}
-\,i\,E\,\Psi_{_{II}}(y_*,z_*)\,i &=& -\,\left[\,\frac{\,\,\,\hbar^{^{2}}}{2\,m}\,\left(\,\partial_{y_*y_*}+\partial_{z_*z_*}\, \right)+ i\, \boldsymbol{h}\cdot \boldsymbol{V}\,\right]\,\Psi_{_{II}}(y_*,z_*) \nonumber \\
 &= &  H\,\,\Psi_{_{II}}(y_*,z_*)\,\,.
\end{eqnarray}
Multiplying the previous equation from the left by the quaternionic conjugate operator of $H$ ($\overline{H}$) and observing that $\overline{H}\,i=i\,H$, we get
\begin{eqnarray*}
E^{^{2}}\,\Psi_{_{II}}(y_*,z_*)  & = &  \overline{H}\,H\,\,\Psi_{_{II}}(y_*,z_*)\\
 & = &\left[\,\frac{\,\,\,\hbar^{^{4}}}{4\,m^{\2}}\,\left(\,\partial_{y_*y_*}+\partial_{z_*z_*}\, \right)^{^{2}} - \frac{\hbar^{^{2}}}{m}\,V_{\1}\, \left(\,\partial_{y_*y_*}+\partial_{z_*z_*}\, \right)   + |\boldsymbol{V}|^{^{2}}\,\right]\,\Psi_{_{II}}(y_*,z_*)\,\,.
\end{eqnarray*}
From the previous equation, we obtain the solution which generalizes the complex solution $q_{z_*}$, i.e.
\begin{equation}
 \label{quat1}
 p_{y_{*}}^{^{2}} + Q_{z_{*}}^{^{2}} = 2\,m\,\left(\,\sqrt{E^{^{2}}-V_{\2}^{^{2}}-V_{\3}^{^{2}}} - V_{\1}\,\right)\,\,,
\end{equation}
which characterizes  the plane wave in the potential region,
\[  \exp[\,i\,(\,p_{y_{*}}\,y_{*}+ Q_{z_{*}}\,z_{*})\,/\hbar\,]\,\,, \]
and the additional solution
\begin{equation}
 \label{quat2}
 p_{y_{*}}^{^{2}} + \widetilde{Q}_{z_{*}}^{^{\,2}} = -\,
 2\,m\,\left(\,\sqrt{E^{^{2}}-V_{\2}^{^{2}}-V_{\3}^{^{2}}} + V_{\1}\,\right)
\end{equation}
which generates {\em evanescent} wave solutions in $z_*$.
We shall give the explicit quaternionic solutions  in the next section. Let us now examine how the Snell law is modified by the {\em new} momentum $Q_{z_*}$. From Eq.(\ref{quat1}), we find
\begin{equation}
\label{pev2}
  p_{y_{*}}^{^{2}} + \,Q_{z_{*}}^{^{2}} = \left(\,\sqrt{1-\frac{V_{\2}^{^{2}}+V_{\3}^{^{2}}}{E^{^{2}}}} - V_{\1}\,\right)\,2\,m\,E =
N^{^2}p^{\2}\,\,.
\end{equation}
The presence of a quaternionic part in our potential generates the refractive index
\begin{equation}  N = \sqrt{\sqrt{1-\frac{V_{\2}^{^{2}}+V_{\3}^{^{2}}}{E^{^{2}}}} - \frac{V_{\1}}{E}}
\end{equation}
and consequently the {\em new} Snell law
\begin{equation}
\sin \theta = N\, \sin \phi\,\,.
\end{equation}
 As it is illustrated in Fig.\,1b, if we replace the complex potential $V_{\1}$ by a pure quaternionic potential of the same modulus $\sqrt{V_{\2}^{^{2}}+V_{\3}^{^{2}}}=E/3$ the transmitted beam will experience a different deflection
\[ \Phi = \arcsin\left[\,\frac{1}{N}\,\sin\theta\,\right] = \arcsin\left[\,\sqrt{\frac{3}{2\sqrt{2}}}\,\frac{1}{\sqrt{2}}\,\right]\approx\frac{\pi}{3.85}\,\,.
\]
The critical angle is now defined in terms of the complex ($V_{\1}$) and quaternionic  ($V_{\2,\3}$) parts of the potential
\begin{equation}
\theta_{_C}\left[\,\frac{V_{\1}}{E}\,,\,\frac{\sqrt{V_{\2}^{^{2}}+V_{\3}^{^{2}}}}{E}\,\right]=
\arcsin\left[\,\sqrt{\sqrt{1-\frac{V_{\2}^{^{2}}+V_{\3}^{^{2}}}{E^{^{2}}}} - \frac{V_{\1}}{E}}\,\right]\,\,,
\end{equation}
which for the case illustrated in Fig.\,1b where $V_{\1}=0$ and $E=3\,\sqrt{V_{\2}^{^{2}}+V_{\3}^{^{2}}}$    gives
\begin{equation}
\theta_{_C}=
\arcsin\left[\,\sqrt{\frac{2\,\sqrt{2}}{3}}\,\right] \approx \frac{\pi}{2.36}\,\,.
\end{equation}
In Fig.\,2, we show the potential dependence for the critical angle. In particular in Fig.\,2a, we compare the critical angle behavior for complex and pure quaternionic potential of the same modulus. The plot clearly show that  pure quaternionic potentials have a greater diffusion zone than complex potentials. This can be explained by noting that
\[ \sin^4\theta_{_{C}}\left[\,0\,,\,x\,\right] - \sin^4\theta_{_{C}}\left[\,x\,,\,0\,\right] =x\,(\,2-x\,)\,\,.  \]
We conclude this section, by observing that for small quaternionic perturbations on complex potential, the {\em new} refractive index $N$ can be rewritten in terms of the standard refractive index $n$ as follows
\[N \,\,\approx \,\, n \,-\, \frac{V_{\2}^{^{2}}+V_{\3}^{^{2}}}{4\,n\,E^{^{2}}}\,\,.\]
The effect of quaternionic perturbations on complex potentials is illustrated in Fig.\,2b.

\section*{\normalsize IV. THE REFLECTION AMPLITUDE FOR QUATERNIONIC POTENTIALS}

To find the reflection amplitude in quaternionic quantum mechanics, we have to impose the continuity
of the wave function and its derivative at the potential discontinuity $d_*$. In the free potential region, $z_*<d_*$, the quaternionic solution is given by\cite{Q1}
\begin{equation}
\Psi_{_{I}}(y_*,z_*) =  \left\{  \exp[i\,p_{z_*}z_*]  + R\, \exp[-\,i\,p_{z_*}z_*] + j\, \widetilde{R}\,\exp[p_{z_*}z_*]   \right\}\,  \exp[i\,p_{y_*}y_*]\,\,.
\end{equation}
In the potential region, $z_*>d_*$, the solution of Eq.(\ref{qse}),  obtained after some algebraic manipulations, reads\cite{Q6,Q7,Q8}
\begin{equation}
\Psi_{_{II}}(y_*,z_*) =  \left\{ (1+j\,\beta)\, T\, \exp[\,i\,Q_{z_*}z_*] + (\alpha +j\,)\,\widetilde{T}\, \exp[\,i\,\widetilde{Q}_{z_*}z_*]   \right\}\,  \exp[i\,p_{y_*}y_*]\,\,,
\end{equation}
where
\[ \alpha= i\,\frac{V_{\2}+i\,V_{\3}}{E+\sqrt{E^{^{2}}-V_{\2}^{^{2}}-V_{\3}^{^{2}}}}\,\,,\,\,\,\,\,\,\,\,\,\,
\beta= -\,i\,\frac{V_{\2}-i\,V_{\3}}{E+\sqrt{E^{^{2}}-V_{\2}^{^{2}}-V_{\3}^{^{2}}}}\,\,,
\]
and
\[
Q_{z_*}=\sqrt{N^{^{2}}-\sin^{\2}\theta}\,\,\,p\,\,,\,\,\,\,\,\,\,\,\,\,
\widetilde{Q}_{z_*}=i\,\sqrt{\sqrt{1- \frac{V_{\2}^{^{2}}+V_{\3}^{^{2}}}{E^{^2}}}+\frac{V_{\1}}{E} +\sin^{\2}\theta }\,\,\,p\,\,.
\]
From the continuity equations, we get
\begin{equation}
\begin{array}{rcl}
\widetilde{R} \exp[p_{z_*}d_*]  & = &   \beta\, T\, \exp[\,i\,Q_{z_*}d_*] + \widetilde{T}\, \exp[\,i\,\widetilde{Q}_{z_*}d_*]\,\,,\\
p_{z_*}\,\widetilde{R} \exp[p_{z_*}d_*]  & = &  i\,\left\{\, \beta\, T\,Q_{z_*}  \exp[\,i\,Q_{z_*}d_*] + \widetilde{T}\,\widetilde{Q}_{z_*} \exp[\,i\,\widetilde{Q}_{z_*}d_*]\, \right\}\,\,,
\end{array}
\end{equation}
which implies
\[ \widetilde{T}\,\exp[\,i\,\widetilde{Q}_{z_*}d_*] =\frac{p_{z_*}-i\,Q_{z_*}}{i\,\widetilde{Q}_{z_*} - p_{z_*}}\,\,\beta\,  T\, \exp[\,i\,Q_{z_*}d_*]\,\,,\]
and
\begin{equation}
\begin{array}{rcl}
 \exp[i\,p_{z_*}d_*]  + R\, \exp[-\,i\,p_{z_*}d_*] & = &   T\, \exp[\,i\,Q_{z_*}d_*] +\alpha\,  \widetilde{T}\, \exp[\,i\,\widetilde{Q}_{z_*}d_*]\\
 & = & \left(\,1 + \alpha\beta\, \displaystyle{\frac{p_{z_*}-i\,Q_{z_*}}{i\,\widetilde{Q}_{z_*} - p_{z_*}}}\,\right)\, T\, \exp[\,i\,Q_{z_*}d_*]\,\,,\\
 \exp[i\,p_{z_*}d_*]  -  R\, \exp[-\,i\,p_{z_*}d_*] & = & \displaystyle{\frac{Q_{z_*}}{p_{z_*}}}\,T\, \exp[\,i\,Q_{z_*}d_*] + \displaystyle{\frac{\widetilde{Q}_{z_*}}{p_{z_*}}}\,\,\alpha\,  \widetilde{T}\, \exp[\,i\,\widetilde{Q}_{z_*}d_*]\\
 & = & \left(\, \displaystyle{\frac{Q_{z_*}}{p_{z_*}}}  + \alpha \beta\, \displaystyle{\frac{\widetilde{Q}_{z_*}}{p_{z_*}}}\,\, \displaystyle{\frac{p_{z_*}-i\,Q_{z_*}}{i\,\widetilde{Q}_{z_*} - p_{z_*}}}\,\right)\, T\, \exp[\,i\,Q_{z_*}d_*]\,\,,
\end{array}
\end{equation}
from which we obtain
\begin{equation}
\label{rquat}
R\left[\frac{V_{\1}}{E}\,,\frac{\sqrt{V_{\2}^{^{2}}+V_{\3}^{^{2}}}}{E}\,;\,\theta\right] =\, \frac{A_{_{-}}}{A_{_{+}}}\,\,\exp[\,2\,i\,p_{z_*}d_*]\,\,,
\end{equation}
where
\[ A_{_\pm}=  (p_{z_*}\pm Q_{z_*})( i\,\widetilde{Q}_{z_*} - p_{z_*})+\alpha\beta\,(p_{z_*}-iQ_{z_*})(p_{z_*} \pm \widetilde{Q}_{z_*})\,\,.  \]
Taking the complex limit of Eq.(\ref{rquat}),  observing that $\alpha\beta \to 0$ and $Q_{z_*}\to \,q_{z_*}$,  we obtain
\begin{eqnarray}
R\left[\frac{V_{\1}}{E}\,,\,0\,;\,\theta\,\right] & = &  \frac{p_{z_*} - q_{z_*}}{p_{z_*} + q_{z_*}}\, \exp[\,2\,i\,p_{z_*}d_*] \nonumber \\
& = & \frac{V_{\1}}{E}\,\exp[\,2\,i\,p_{z_*}d_*]\,\,\mbox{\huge $/$} \left(\,\cos\theta + \sqrt{\cos^{\2}\theta -\displaystyle{\frac{V_{\1}}{E}}}\,\right)^{^{2}}
\end{eqnarray}
and recover the result given in section II, see  Eq.(\ref{rcom}). Observe that for $E\,\cos\theta<V_{\1}$ we have total internal reflection.  In Fig.\,3, we plot the reflection amplitude behavior for fixed incidence angles
as a function of the potential (see Fig.\,3a) and for a fixed potential as a function of the incidence angle (see Fig\,3b). From the plots it is evident that by increasing the quaternionic part of the potential we increase the diffusion zone.

\section*{\normalsize V. CONCLUSIONS}

The well-known analogy between optics and complex quantum mechanics\cite{DRAG,SL} allow to obtain the relationship between the sine of the angles of incidence, $\theta$, and refraction, $\varphi$, of waves passing through a boundary between two different isotropic media, i.e
\[ \sin\theta = n\,\sin \psi\,\,,\]
by analyzing the planar motion of a quantum particle  moving from the free region, $z_*<d_*$,
to the potential region $V_{\1}$, $z_*>d_*$.  In the complex case the refractive index is given by
\[ n = \sqrt{1-\frac{V_{\1}}{E}}\,\,,\]
where $E$ is the energy of the incoming particle. In this paper, we have extended such a study
to quaternionic potentials. The presence of a quaternionic part in the potential modified the refractive index as follows
 \[ N = \sqrt{\sqrt{1-\frac{V_{\2}^{^{2}}+V_{\3}^{^{2}}}{E^{^{2}}}} - \frac{V_{\1}}{E}}\,\,.\]
Consequently, the  {\em new} Snell law becomes
\[
\sin \theta = N\, \sin \phi\,\,.
\]
The potential dependence of the critical angle, see Fig.\,2, and the reflection amplitude behavior as a function of the complex/quaternionic potential ratio, see Fig.\,3a, and of the incidence angle, see Fig.\,3b, allow to know in which situations we can distinguish between  complex and quaternionic potentials.

Finally, in the tunneling energy zone,  $\sqrt{V_{\2}^{^{2}}+V_{\3}^{^{2}}}<E$ and $|\boldsymbol{V}|>E$, we find $Q_{z_*} = i\, |Q_{z_*}|$,   $\widetilde{Q}_{z_*} = i\, |\widetilde{Q}_{z_*}|$ and $\alpha\beta \in \mathbb{R}$. In this limit,
\[ A_{_\pm} \,\to\,  -\,(p_{z_*}\pm i\,|Q_{z_*}|)(p_{z_*} + |\widetilde{Q}_{z_*}|  )+\alpha\beta\,(p_{z_*}+|Q_{z_*}|)(p_{z_*} \pm i\,|\widetilde{Q}_{z_*}|)\,\,\,\Rightarrow\,\,\,\left|R\right|=1\,\,. \]
For total reflection, the addition phase in the reflection coefficient is responsible for the
Goos-H\"anchen shift\cite{GH}. In a forthcoming paper, we shall investigate in detail the phase change which takes place on total reflection. This study could be useful to identify in which dielectric systems quaternionic deviations from the complex
optical path can be seen. Another interesting generalization of the analysis presented in this paper is represented by the possibility to extend the non relativistic discussion to the quaternionic relativistic case\cite{Sou1,Sou2} and eventually to Clifford algerbas\cite{Cli1,Cli2,Cli3}.

\newpage

\begin{figure}
\vspace*{-2cm} \hspace*{-2cm}
\includegraphics[width=19cm, height=25cm, angle=0]{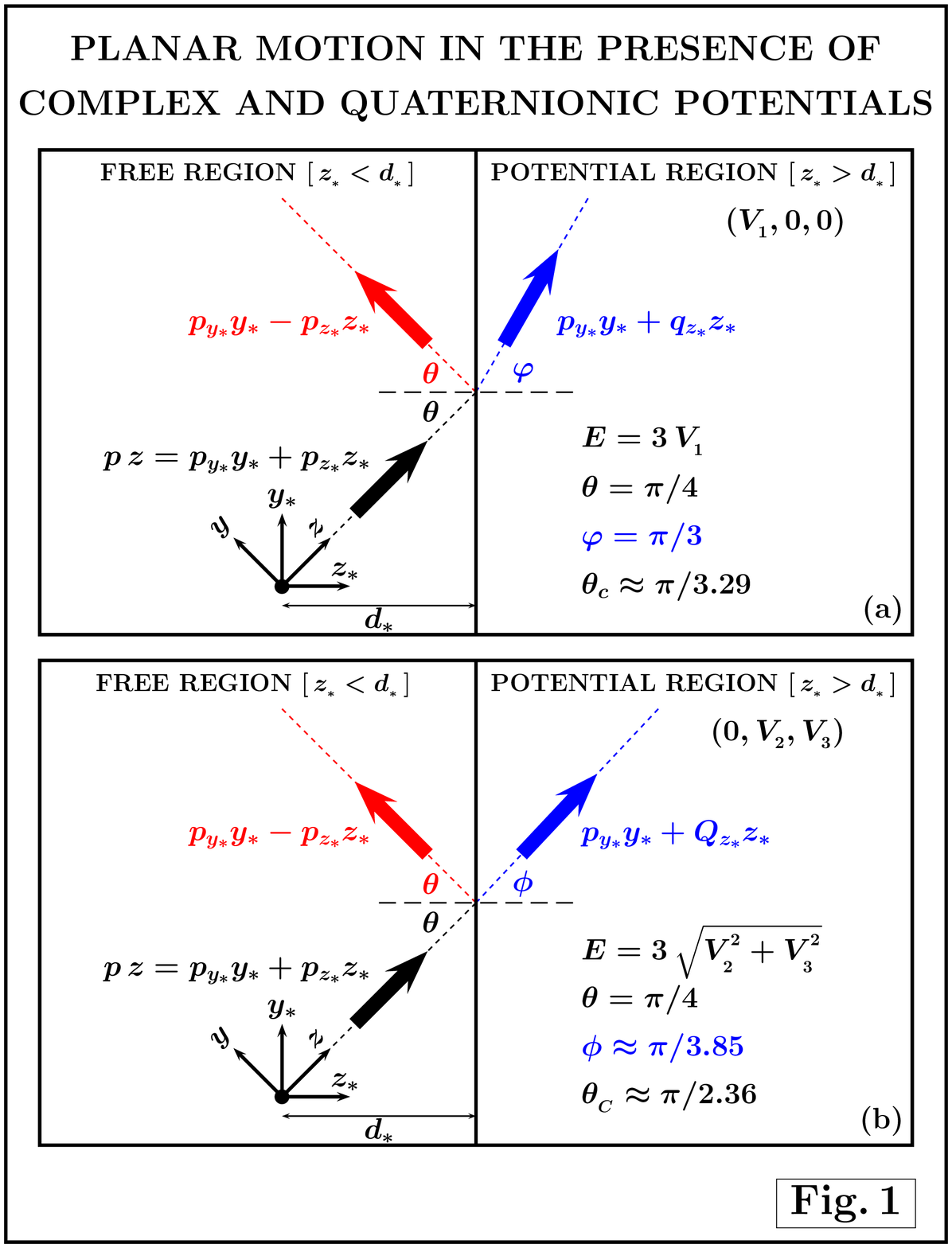}
\vspace*{-2.5cm}
 \caption{Planar motion in the presence of complex (a) and pure quaternionic (b) potentials. For a given incidence angle, complex potentials present a deflection greater than pure quaternionic potentials. As a consequence the critical angle is greater for pure quaternionic than for complex potentials.}
\end{figure}
\newpage

\begin{figure}
\vspace*{-2cm} \hspace*{-2cm}
\includegraphics[width=19cm, height=25cm, angle=0]{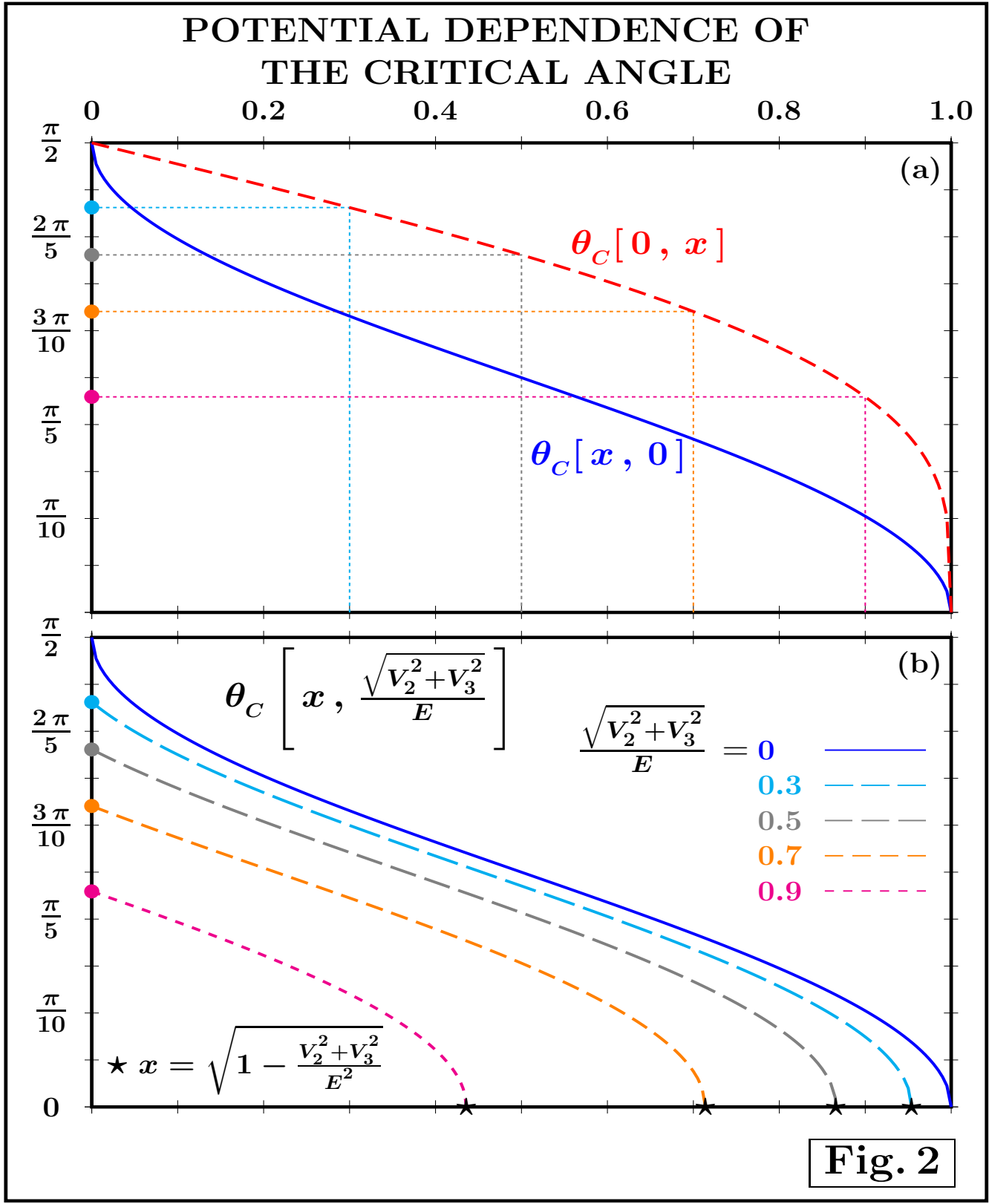}
\vspace*{-2.5cm}
 \caption{Potential dependence of the critical angle  $\theta_{_C}[\,V_{\1}/E\,,\, \sqrt{V_{\2}^{^{2}}+V_{\3}^{^{2}}}/E\,]$. In (a), we compare complex with pure quaternionic potentials  of the same modulus. In (b), we illustrated how the introduction of quaternionic perturbations on complex potentials modifies the critical angle.}
\end{figure}

\newpage
\begin{figure}
\vspace*{-2cm} \hspace*{-2cm}
\includegraphics[width=19cm, height=25cm, angle=0]{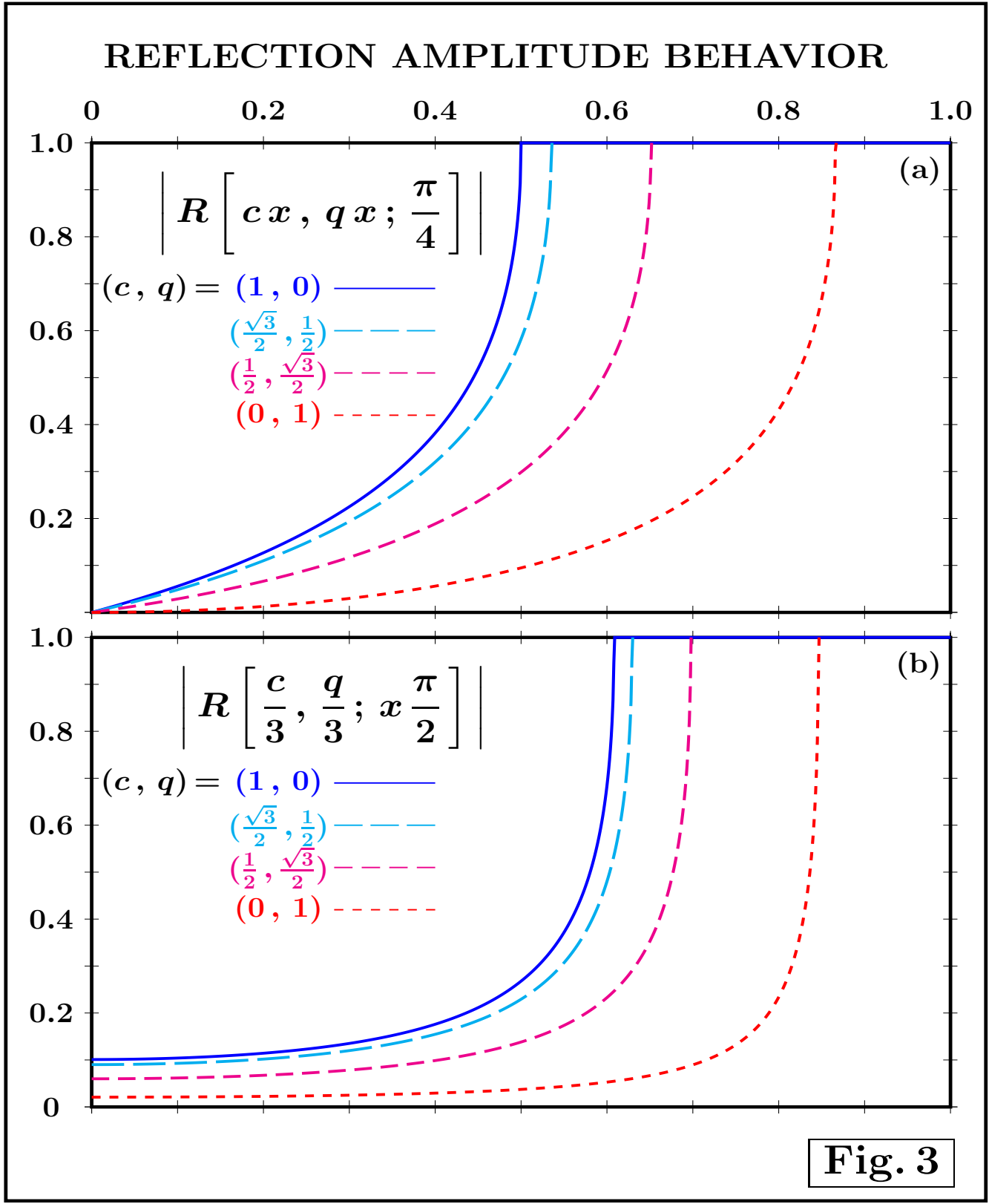}
\vspace*{-2.5cm}
 \caption{Reflection amplitude behavior for a fixed incidence angle as function of the potential (a) and
 for a fixed potential as function of the incidence angle (b). The plots clearly show that, by increasing the quaternionic part of the potential, we increase the diffusion zone. Consequently the reflection probability decreases with respect to complex potentials.}
\end{figure}

\end{document}